\documentclass[dvips,rmp,twocolumn,groupedaddress,floatfix]{revtex4}

\makeatletter
\renewcommand\@biblabel[1]{#1.}
\makeatother
\bibpunct{(}{)}{,}{a}{,}{,}
\usepackage[dvips]{graphicx}
\usepackage[T1]{fontenc}
\usepackage{bm}
\usepackage{dcolumn,graphicx,amsmath,amssymb,txfonts}
\newcommand{\argmax}{\operatornamewithlimits{max}}
\newcommand{\T}{\rule{0pt}{2.6ex}}

\begin{document}


\title{An information-theoretic framework for resolving community structure in complex networks}

\author{Martin Rosvall}
\email{rosvall@u.washington.edu}
\homepage{http://www.tp.umu.se/~rosvall/}
\affiliation{Department of Biology, University of Washington, Seattle, WA 98195-1800}
\author{Carl T. Bergstrom}
\homepage{http://octavia.zoology.washington.edu/}
\affiliation{Department of Biology, University of Washington, Seattle, WA 98195-1800}

\date{\today}

\begin{abstract}
To understand the structure of a large-scale biological, social, or technological network, it can be helpful to decompose the network into smaller subunits or modules. In this article, we develop an information-theoretic foundation for the concept of modularity in networks. We identify the modules of which the network is composed by finding an optimal compression of its topology, capitalizing on regularities in its structure. We explain the advantages of this approach and illustrate them by partitioning a number of real-world and model networks.
\end{abstract}

\maketitle

Many objects in nature, from proteins to humans, interact in groups that compose social \citep{freeman}, technological \citep{eriksen}, or biological systems \citep{hartwell}.
The groups form a distinct intermediate level between the microscopic and macroscopic descriptions of the system, and group structure may often be coupled to aspects of system function including robustness \citep{hartwell} and stability \citep{variano}. When we map the interactions among components of a complex system to a network with nodes connected by links, these groups of interacting objects form highly connected modules that are only weakly connected to one other. We can therefore comprehend the structure of a dauntingly complex network by identifying the modules or communities of which it is composed \citep{girvan_newman,palla,guimera-nature,holme,newman-rew2004,danon2005}.
When we describe a network as a set of interconnected modules, we are highlighting certain regularities of the network's structure while filtering out the relatively unimportant details. Thus a modular description of a network can be viewed as a lossy compression of that network's topology, and the problem of community identification as a problem of finding an efficient compression of the structure.

This view suggests that we can approach the challenge of identifying the community structure of a complex network as a fundamental problem in information theory \citep{shannon,rissanen1978,ziv}.
We provide the groundwork for an information-theoretic approach to community detection, and explore the advantages of this approach relative to other methods for community detection.

\begin{figure*}
\centering
\includegraphics[width=1.0\textwidth]{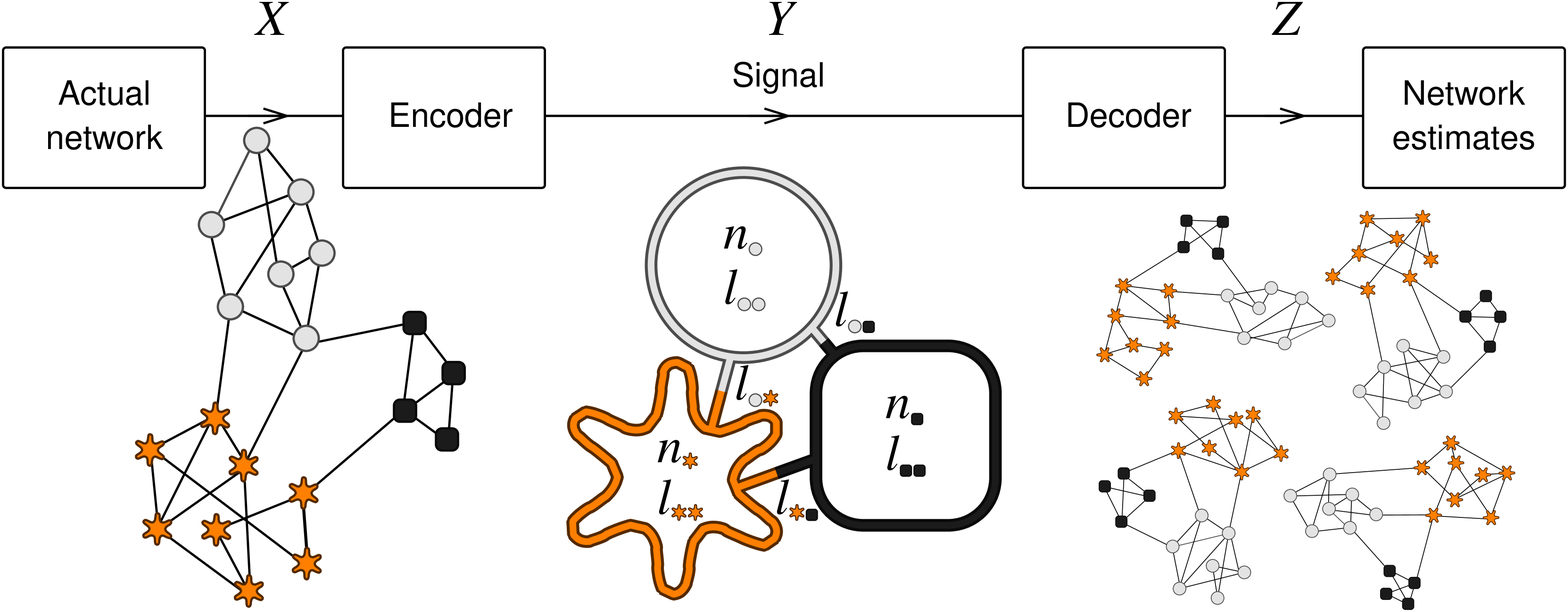}
\caption{\label{fig1}Basic framework for detecting communities as a communication process. A signaler knows the full network  structure and wants to send as much information as possible about the network to a receiver over a channel with limited capacity. The signaler therefore encodes the network into modules in a way that maximizes the amount of information about the original network. This figure illustrates an encoder that compresses the network into 3 modules
$i=$
{\protect \includegraphics[width=1.7ex]{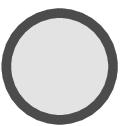}},
{\protect \includegraphics[width=1.7ex]{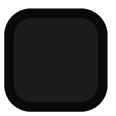}},
{\protect \includegraphics[width=1.7ex]{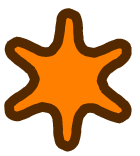}},
with $n_i$ nodes and $l_{ii}$ links in each module, and $l_{ij}$ links between the modules. The receiver can then decode the message and construct a set of possible candidates for the original network. The smaller the set of candidates, the more information the signaler has managed to transfer.}
\end{figure*}

Figure \ref{fig1} illustrates our basic framework for identifying communities. We envision the process of describing a complex network by a simplified summary of its module structure as a communication process. The link structure of a complex network is a random variable $X$;  a signaler knows the full form of the network $X$,  and aims to convey much of this information in a reduced fashion to a signal receiver. To do so, the signaler encodes information about $X$ as some simplified description $Y$. She sends the encoded message through a noiseless communication channel. The signal receiver observes the message $Y$, and then ``decodes'' this message, using it to make guesses $Z$ about the structure of the original network $X$.

There are many different ways to describe a network $X$ by a simpler description $Y$. Which of these is best? The answer to this question of course depends on what you want to do with the description. Nonetheless, information theory offers an appealing general answer to this question. Given some set of candidate descriptions $Y_i$, the best description $Y$ of a random variable $X$ is the one that tells the most about $X$ --- that is, the one that maximizes the mutual information $I(X;Y)$ between description and network.

Since we are interested in identifying community structure, we will explore descriptions $Y$ that summarize the structure of a network $X$ by enumerating the communities or modules within $X$, and describing the relations among them. In this paper, we will consider one particular method of encoding the community structure of $X$. More generally one could and indeed should consider alternative ``encoders,'' so as to choose one best suited for the problem at hand.

We consider an unweighted and undirected network $X$ of size $n$ with $l$ links, which can be described by the adjacency matrix 
\begin{equation}
\mathbf{A}_{ij} = \begin{cases}
           \enspace 1 & \text{if there is a link between nodes $i$ and $j$} \\
           \enspace 0 & \text{otherwise.}
         \end{cases}
\end{equation}
We choose the description 
\begin{equation}\label{descriptionY}
 Y = \left\{ \begin{array}{c}
\mathbf{a}=\left(\begin{array}{c}
a_1\\
\vdots\\
\vdots\\
a_n
\end{array}\right),
\;\mathbf{M}=
\left(\begin{array}{ccc}
l_{11} & \cdots & l_{1m}\\
\vdots & \ddots & \vdots\\
l_{m1} & \cdots & l_{mm}
\end{array}\right)
\end{array}\right\}
\end{equation}
for $m$ modules, where $\mathbf{a}$ is the module assignment vector, $a_i \in \{1,2,\ldots,m\}$, and $\mathbf{M}$ is the module matrix.
The module matrix $\mathbf{M}=\mathbf{M}(X,\mathbf{a})$ describes how the $m$ modules given by the assignment vector are connected in the actual network. Module $i$ has $n_i$ nodes and connects to module $j$ with $l_{ij}$ links (see Fig.~\ref{fig1}).

To find the best assignment $\mathbf{a}^{*}$ we now maximize the mutual information over all possible assignments of the nodes into $m$ modules
\begin{equation}
 \mathbf{a}^{*} = \arg\argmax_{\mathbf{a}} I(X;Y).
\end{equation}
By definition, the mutual information $I(X;Y)=H(X)-H(X|Y)=H(X)-H(Z)$, where $H(X)$ is the information necessary to describe $X$ and the conditional information $H(X|Y)=H(Z)$ is the information necessary to describe $X$ given $Y$ (see Fig.~\ref{fig1}). We therefore seek to to minimize $H(Z)$. This is equivalent to constructing an assignment vector such that the set of network estimates $Z$ in Fig.~\ref{fig1} is as small as possible. Given that the description $Y$ assigns nodes to $m$ modules,
\begin{equation}\label{zentropy}
 H(Z) = \log\left[ \prod_{i=1}^{m}\binom{n_i(n_i-1)/2}{l_{ii}}\prod_{i > j}\binom{n_in_j}{l_{ij}} \right],
\end{equation}
where the parentheses denote the binomial coefficients and the logarithm is taken in base 2. Each of the $m$ binomial coefficients in the first product gives the number of different modules that can be constructed with $n_i$ nodes and $l_{ii}$ links. Each of the $m(m-1)/2$ binomial coefficients in the second product gives the number of different ways module $i$ and $j$ can be connected to one another.\\

\begin{figure}
\centering
\includegraphics[width=1.0\columnwidth]{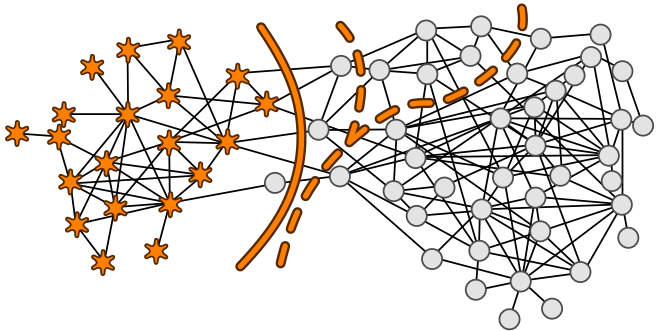}
\caption{\label{fig2}The dolphin network by \citep{lusseau} partitioned with our cluster-based compression (solid line) and based on the modularity (dashed line). The stars and circles represent the two observed groups of dolphins. The right branch of the dashed line represents a split based on maximizing the modularity, which is different from the left branch solution based on the spectral analysis approximation presented in ref.\ \citep{newman-2006-74}. The edge-betweenness algorithm presented in ref.~\citep{girvan_newman} splits the network in the same way as our cluster-based compression method \citep{newmangirvan2004}.}
\end{figure}

In Fig.~\ref{fig2} we apply our cluster-based compression method to the dolphin social network reported by Lusseau \textit{et al}.\ \citep{lusseau}. Our method selects a division that differs by only one node from the division along which the actual dolphin groups were observed to split. 
Because it is computationally infeasible to check all possible partitions of even modestly-sized networks, we use 
simulated annealing with the heat-bath algorithm to search for the partition that maximizes the mutual information between the description and the original network. We have confirmed the results for the networks in the figures with exhaustive searches in the vicinity of the Monte Carlo solutions.

We compare our results with the partition obtained by using the modularity approach introduced by Newman and Girvan in ref.\ \citep{newmangirvan2004}; that technique has been widely adopted because of its appealing simplicity, its performance in benchmark tests \citep{danon2005}, and the availability of powerful numerical techniques for dealing with large networks \citep{newman-fast,newman-spectral,reichardt:016110}. 
Given a partitioning into $m$ modules, the modularity $Q$ is the sum of the contributions from each module $i$
\begin{equation}\label{moularity}
 Q = \sum_{i=1}^{m} l_{ii}/l-(d_i/2l\,)^2,
\end{equation}
where $l_{ii}$ is the number of links between nodes in the $i$-th module, $d_i$ the total degree in module $i$,  and $l$ is the total number of links in the network. When we maximize the modularity, we are not just minimizing the number of links between modules. Instead we find a configuration which maximizes the number of links within modules in the actual network, minus the expected number of links within comparable modules in a random network with the same degree sequence \citep{newman-spectral}. Or equivalently, we aim to divide the network such that the number of links within modules is higher than expected.

This approach works beautifully for networks where the modules are similar in size and degree sequence \citep{danon2005}. However, when the dolphin network in Fig.~\ref{fig2} is partitioned using the modularity approach, the network ends up being divided very differently from the empirically observed fission of the dolphin group. Why? Because of the $(2l\,)^2$ denominator in the second term of the definition of modularity (Eq.~\ref{moularity}), the choice of partition is highly sensitive to the total number of links in the system. By construction, the benefit function that defines modularity favors groups with similar total degree, which means that the size of a module depends on the size of the whole network \citep{fortunato}.

To compare quantitatively the performance of our cluster-based compression method with modularity-based approaches, we conducted the benchmark tests described in refs.\ \citep{girvan_newman,danon2005}. In these tests, 128 nodes are divided into four equally sized groups with average degree 16. As the average number of links $k_{\mathrm{out}}=6,7,8$ from each node to nodes in other groups increases, it becomes harder and harder to identify the underlying group structure. 

Table \ref{table} presents the results from both methods using the simulated-annealing scheme described above for the numerical search; we obtained comparable results for networks with up to $10^4$ nodes. When the groups are of equal size and similar total degree, both methods perform very well, on par with the best results reported in refs.\ \citep{guimera-nature,guimera-jstat}.

When the groups vary in size or in total degree, as was the case in the dolphin network, the modularity approach has more difficulty resolving the community structure (Table \ref{table}).

We merged three of the four groups in the benchmark test to form a series of test networks each 
with one large group of 96 nodes and one small group with 32 nodes. These asymmetrically-sized networks are harder for either approach to resolve, but cluster-based compression recovers the underlying community structure more often than does modularity, by a sizable margin. Finally we conducted a set of benchmark tests using networks composed of two groups each with 64 nodes, but with different average degrees of 8 and 24 links per node. For these networks, we use $k_{\mathrm{out}}=2,3,4$, and cluster-based compression again recovers community structure more often than does modularity.\\

\begin{table}[pbt]
\centering
\begin{minipage}{1.0\columnwidth}
\centering
\caption{\label{table} Benchmark performance for symmetric and asymmetric group detection measured as fraction of correct assignments, averaged over 100 network realizations with the standard deviation in the parentheses. Here we assume that the true number of modules (4 in the symmetric case, 2 in the asymmetric cases) is known a priori. 
}
\begin{tabular*}{1.0\columnwidth}{@{\extracolsep{\fill}}lrrr}
\hline
\T Symmetric & \multicolumn{1}{c}{$k_{\mathrm{out}}=6$} & \multicolumn{1}{c}{$k_{\mathrm{out}}=7$} & 
\multicolumn{1}{c}{$k_{\mathrm{out}}=8$}\\
\rule{2ex}{0pt}\textit{Compression} & 0.99 (.01) & 0.97 (.02) &  0.87 (.08)\\
\rule{2ex}{0pt}\textit{Modularity} & 0.99 (.01) & 0.97 (.02) & 0.89 (.05)\\
\T Node asymmetric & \multicolumn{1}{c}{$k_{\mathrm{out}}=6$} & \multicolumn{1}{c}{$k_{\mathrm{out}}=7$} & 
\multicolumn{1}{c}{$k_{\mathrm{out}}=8$}\\
\rule{2ex}{0pt}\textit{Compression} & 0.99 (.01) & 0.96 (.04) &  0.82 (.10)\\
\rule{2ex}{0pt}\textit{Modularity} & 0.85 (.04) & 0.80 (.03) & 0.74 (.05)\\
\T Link asymmetric & \multicolumn{1}{c}{$k_{\mathrm{out}}=2$} & \multicolumn{1}{c}{$k_{\mathrm{out}}=3$} & \multicolumn{1}{c}{$k_{\mathrm{out}}=4$}\\
\rule{2ex}{0pt}\textit{Compression}\footnote{The standard deviation is non-zero but less than 0.01 for $k_{\mathrm{out}}=2,3$.} & 1.00 (.00) & 1.00 (.00) &  1.00 (.01)\\
\rule{2ex}{0pt}\textit{Modularity} & 1.00 (.01) & 0.96 (.03) & 0.74 (.10)\\
\hline
\end{tabular*}
\begin{footnotesize}
\end{footnotesize}
\end{minipage}
\end{table} 

\begin{figure}[tpb]
\centering
\includegraphics[width=1.0\columnwidth]{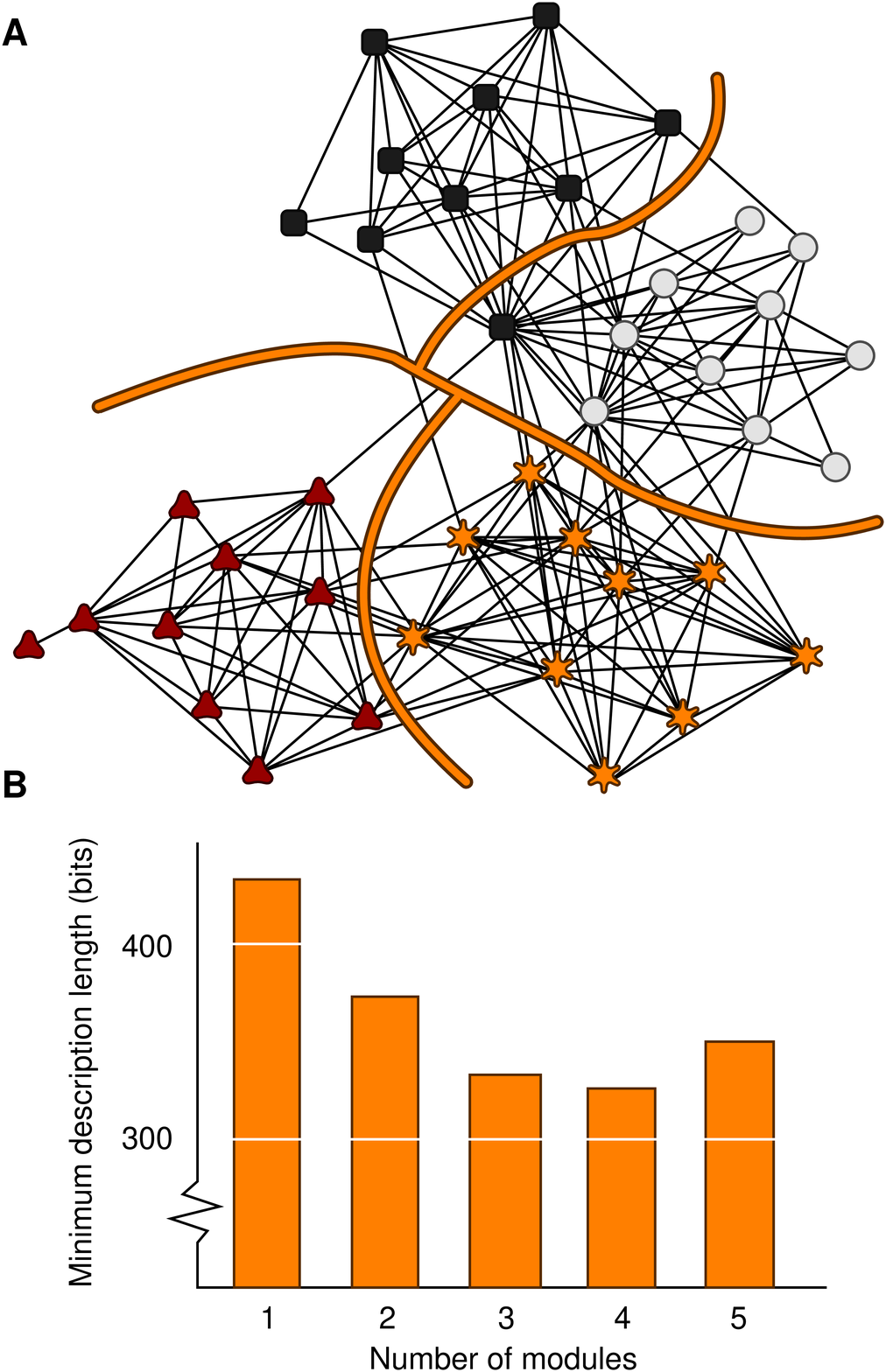}
\caption{\label{fig3} Partitioning into an optimal number of modules. The network in panel A consists of 40 journals as nodes from four different fields: multidisciplinary physics (squares), chemistry (circles), biology (stars), and ecology (triangles). The 189 links connect nodes if at least one article from one of the journals cites an article in the other journal during 2004 {\protect\citep{jsr}}. We have selected the 10 journals with the highest impact factor in the four different fields, but disregarded journals classified in one or more of the other fields. Panel B shows the minimum description length for the network in panel A partitioned into 1 to 5 different modules. The optimal partitioning into four modules is illustrated by the lines in panel A.}
\end{figure}

Next, we address a model selection challenge. In some special cases we will know a priori how many modules compose our sample network, but in general the task of resolving community structure is twofold. We must determine the number of modules in the network, and then we need to partition the nodes into that number of modules. The catch is that we cannot determine the optimal number of modules without also considering the assignments of nodes --- so these problems need to be solved simultaneously.  Below, we provide a solution grounded in algorithmic information theory.

Looking back at Fig.~\ref{fig1}, the encoder seeks to find a compression of the network so that the decoder can make the best possible estimate of the actual network. One approach would be to have the encoder partition the network into $n$ modules, one for each node. This ensures that the decoder can reconstruct the network completely, but under this approach nothing is gained either in compression or module identification. Therefore the encoder must balance the amount of information necessary to describe the network in modular form, as given by the signal $Y$ in Fig.~\ref{fig1}, and the uncertainty that remains once the decoder receives the modular description, as given by the size of the set of network estimates $Z$ in Fig.~\ref{fig1}. This is an optimal coding problem and can be resolved by the Minimum Description Length (MDL) principle \citep{rissanen1978,barron,grunwald}. The idea is to exploit the regularities in the structure of the actual network $X$ to summarize it in condensed form, without overfitting it. What do we mean by overfitting in this context? Figure \ref{fig3} illustrates. We want to choose a set of modules for the journal citation network in Fig.~\ref{fig3} such that if we were to repeat the experiment next year, each journal would likely be assigned to the same module again. If we overfit the data, we may capture more of a specific year's data, but unwittingly we also capture noise that will not recur in next year's data. 

To minimize the description length of the original network $X$, we look for the number of modules $m$ that minimizes that the length of the modular description $Y$ plus the ``conditional description length'', where the conditional description length is the amount of additional information that would be needed to specify $X$ exactly to a receiver who had already decoded the description $Y$ \citep{rissanen1978}. That is, we seek to minimize the sum
\begin{equation}
 L(Y) + L(X | Y),
\end{equation}
where $L(Y)$ is the length in bits of the signal and $L(X | Y)$ is number of bits needed to specify which of the network estimates implied by the signal $Y$ is actually realized. The description length is easy to calculate in this discrete case and is given by
\begin{equation}
 L(Y) + L(X | Y) = n\log{m}+\frac{1}{2}m(m+1)\log{l}+H(Z),
\end{equation}
where the first and second term give the size necessary to encode the assignment vector $\mathbf{a}$ and the module matrix $\mathbf{M}(X,\mathbf{a})$, and $H(Z)$ is given in Eq.~\ref{zentropy}. Figure \ref{fig3}B shows the description length with the journal network partitioned into one to five modules. Four modules yield the minimum description length and we show the corresponding partition in Fig.~\ref{fig3}A.

This cluster-based compression assigns 39 of the 40 journals into the proper categories, but places the central hub \emph{Physical Review Letters} (PRL) in the chemistry cluster. This may seem like a mistake, given that PRL has 9 links to physics and only 8 to chemistry. Indeed, a partitioning based on the modularity score $Q$ places PRL among the physics journals. But whatever its subject matter, the structural role that PRL plays in the unweighted journal network is that of a chemistry journal.
Like most of the chemistry journals, and unlike its compatriots in physics, PRL is closely linked to biology and somewhat connected to ecology.

We can also partition the network into two, three, or five modules, but doing so yields a longer total description length. When we compress the network into two components, physics clusters together with chemistry and biology clusters together with ecology. When we split into three components, ecology and biology separate but physics and chemistry remain together in a single module. When we try to split the network into five modules, we get essentially the same partition as with four, only with the singly connected journal \emph{Conservation Biology} split off by itself into its own partition. One might not even consider that singleton to be a valid module. 

To get a sense of how different methods handle the model selection problem, we  compared the performance of our cluster-based compression method with the modularity-based approach. Instead of looking for the best assignment given the correct number of modules as in Table \ref{table}, we look at the performance of each method at estimating the correct number of modules. Our results are summarized in Table \ref{table2}. Both cluster-based compression and modularity exhibit thresholds beyond which they are unable with high probability to reconstruct the underlying module structure that generated the data.  Beyond this threshold, the compression method tends to underestimate the number of groups. By contrast, the modularity tends to overestimate the number of groups. Others have observed similar model selection bias by the modularity approach; in a completely random network, the modularity-based approach typically detects multiple and therefore statistically insignificant modules \citep{guimera-pre,reichardt-physD,alon}.

When the clusters are symmetric in size and degree, both methods reach the resolution threshold at approximately the same point. However, when the groups have unequal numbers of nodes or unequal degree distributions, the cluster-based compression method is able to successfully reconstruct the underlying structure of networks that the modularity approach cannot recover (Table \ref{table2}).\\

\begin{table}[pbt]
\centering
\begin{minipage}{1.0\columnwidth}
\centering
\caption{\label{table2} Benchmark performance for model selection measured as fraction of correct identification of number of groups, averaged over 100 network realization with the average number of assigned modules in the parentheses}
\begin{tabular*}{1.0\columnwidth}{@{\extracolsep{\fill}}lrrr}
\hline
\T Symmetric\footnote[1]{The symmetric networks have four modules.} & \multicolumn{1}{c}{$k_{\mathrm{out}}=6$} & \multicolumn{1}{c}{$k_{\mathrm{out}}=7$} & 
\multicolumn{1}{c}{$k_{\mathrm{out}}=8$}\\
\rule{2ex}{0pt}\textit{Compression} & 1.00 (4.00) & 1.00 (4.00) &  0.14 (1.93)\\
\rule{2ex}{0pt}\textit{Modularity} & 1.00 (4.00) & 1.00 (4.00) & 0.70 (4.33)\\
\T Node asymmetric\footnote[2]{The asymmetric networks have two modules.} & \multicolumn{1}{c}{$k_{\mathrm{out}}=6$} & \multicolumn{1}{c}{$k_{\mathrm{out}}=7$} & 
\multicolumn{1}{c}{$k_{\mathrm{out}}=8$}\\
\rule{2ex}{0pt}\textit{Compression} & 1.00 (2.00) & 0.80 (1.80) & 0.06 (1.06)\\
\rule{2ex}{0pt}\textit{Modularity} & 0.00 (4.95) & 0.00 (4.97) & 0.00 (5.29)\\
\T Link asymmetric${}^{b}$ & \multicolumn{1}{c}{$k_{\mathrm{out}}=2$} & \multicolumn{1}{c}{$k_{\mathrm{out}}=3$} & \multicolumn{1}{c}{$k_{\mathrm{out}}=4$}\\
\rule{2ex}{0pt}\textit{Compression} & 1.00 (2.00) & 1.00 (2.00) & 1.00 (2.00)\\
\rule{2ex}{0pt}\textit{Modularity} & 0.00 (3.10) & 0.00 (4.48) & 0.00 (5.55)\\
\hline
\end{tabular*}
\end{minipage}
\end{table} 

Let us look back at the journal network in Fig.~\ref{fig3}, and recall that we cannot partition this network into more than four modules without creating at least one module that has a majority of its links to nodes in other modules. Because of this concept of what a module is \citep{wasserman,radicchi}, we might be interested only in those clusters with more links within than between clusters ($l_{ii}>l_{ij}\text{ for all }i \text{ and } j$ in Eq.~\ref{descriptionY}). However, choosing modules in that way will not necessarily maximize mutual information. In many cases we get a higher mutual information by selecting modules such that hubs are clustered together and peripheral nodes are clustered together. When this is true, we can describe the network structure more efficiently by clustering nodes with similar roles instead of clustering nodes that are closely connected to one another. The mixture model approach provides an alternative method of identifying aspects of network structure beyond positive assortment \citep{newman-mixture}. In our examples where we want to find modules with more links within modules than between them, we impose a ``link constraint'', penalizing solutions with more links between than within in the simulated annealing scheme.

\begin{figure}[tpb]
\centering
\includegraphics[width=0.9\columnwidth]{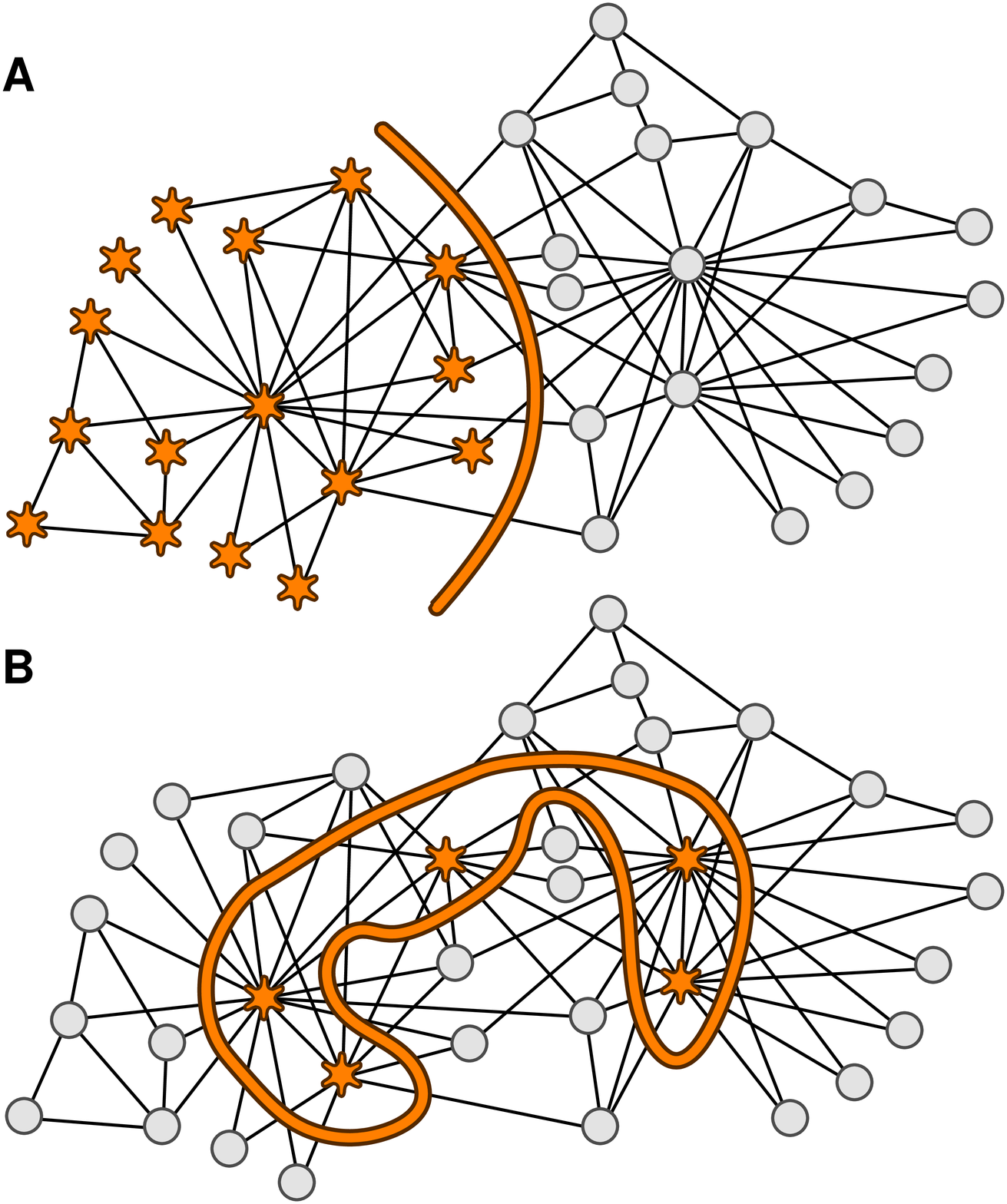}
\caption{\label{fig4} Zachary's karate club network \citep{karateclub} partitioned into two modules based on the maximum mutual information with (panel A) and without (panel B) the link constraint. The partitioning with more links within modules than between modules in panel A clusters closely connected nodes together and the unconstrained partitioning in panel B clusters nodes with similar roles together.}
\end{figure}

To visualize the different ways of partitioning a network, we split Zachary's classic karate club network \citep{karateclub} with (panel A) and without (panel B) the link constraint (Fig.~\ref{fig4}). In panel A the partitioning corresponds exactly to the empirical outcome that was observed by Zachary, but in panel B the 5 members with the highest degrees are clustered together. In the first case the compression capitalizes on the high frequency of ties between members of the same subgroup and the relatively few connections between the groups. In the second case the compression takes advantage of the very high number of links between the five largest hubs and the peripheral members, and the very few connections between the peripheral members. The compression with the hubs in one cluster and the peripheral nodes in the other cluster is in this case more efficient.

\section*{Conclusions}

We have shown that the process of resolving community structure in complex networks can be viewed as a problem in data compression. By drawing out the relationship between module detection and optimal coding we are able to ground the concept of network modularity in the rigorous formalism provided by information theory. 

Enumerating the modules in a network is an act of description; there is an inevitable tradeoff between capturing most of the network structure at the expense of needing a long description with many modules, and omitting some aspects of network structure so as to allow a shorter description with fewer modules. Our information-theoretic approach suggests that there is a natural scale on which to describe the network, thereby balancing this tradeoff between under- and over-description.

The main purpose of this manuscript is to propose a new theoretical approach to community detection, and thus we have not extensively explored methods of optimizing the computational search procedure. 
Nonetheless, we have partitioned networks of sizes up to $10^4$ nodes with a simple simulated annealing approach.
While many interesting real-world networks are smaller than this, it is our hope that the approach can be used for even larger networks with other optimization methods such as the greedy search technique presented in ref.~\citep{clauset-2004-70}.

For many networks, our cluster-based compression method yields somewhat different results than does the modularity approach developed by Newman and colleagues.
The differences reflect alternative perspectives on what community structure might be. If one views community structure as statistical deviation from the null model in which the degree sequence is held constant but links are otherwise equiprobable among all nodes, the modularity optimization method by definition provides the optimal partitioning. If one views community structure as the regularities in a network's topology that allow the greatest compression of the network's structure, our approach provides a useful partitioning. The choice of which to pursue will depend on the questions that a researcher wishes to ask.

In this paper, we have concentrated on finding communities of
nodes that are positively clustered by the links among them. While this
is a common goal in community detection, the sort of information
that we wish to extract about network topology may vary from application
to application. By choosing an appropriate encoder, one can identify other
aspects of structure, such as hub versus periphery distinction
illustrated in our alternative partitioning of the karate club network.
When we abstract the problem of finding pattern in networks to a problem
of data compression, the information-theoretic
view described here provides a general basis for how to get the most
information out of a network structure.

\begin{acknowledgments}
We thank Ben Althouse for generating the network used in Fig.~\ref{fig3} and Mark Newman for constructive comments on the manuscript. This work was supported by the National Institute of General Medical Sciences
Models of Infectious Disease Agent Study program cooperative agreement 5U01GM07649.
\end{acknowledgments}

\end{document}